\documentclass[letterpaper,12pt]{article} 
\usepackage{amsmath,amsfonts}
\usepackage{latexsym}
\providecommand{\e}[1]{\mspace{2mu}\mathrm{e}^{#1}\mspace{1mu}}
\providecommand{\sgn}[1]{\mspace{2mu}\mathrm{sgn}\mspace{2mu}#1\,}
\providecommand{\I}{\mathrm{I}}
\providecommand{\K}{\mathrm{K}}
\providecommand{\J}{\mathrm{J}}
\providecommand{\N}{\mathrm{N}}
\providecommand{\mpl}[1]{M_{Pl(#1)}}

\title{Consistent Linearized Gravity in Brane Backgrounds}
\author{I.~Ya.~Aref'eva$^b$, M.~G.~Ivanov$^{c,d}$, W.~M\"uck$^a$,\\[6pt]
K.~S.~Viswanathan$^a$ and I.~V.~Volovich$^{b,a}$\thanks{E-mail addresses:
\texttt{arefeva@mi.ras.ru, mgi@socrates.berkeley.edu or mgi@mi.ras.ru,
wmueck@sfu.ca, kviswana@sfu.ca, volovich@mi.ras.ru}} \\[6pt]
\emph{\small$^a$ Department of Physics, Simon Fraser University,}\\ 
\emph{\small Burnaby, B.C., V5A 1S6 Canada} \\
\emph{\small$^b$ Steklov Mathematical Institute, Gubkin St.~8, 117966
Moscow, Russia}\\
\emph{\small$^c$ Department of Physics, University of California,
94720--7300 Berkeley, USA}\\
\emph{\small$^d$ Moscow Institute of Physics and Technology,}\\ 
\emph{\small Institutsky Per.~9, Dolgoprudny, Moscow Reg., Russia} }
\date{}

\begin{document}

\maketitle
\abstract{A globally consistent treatment of linearized gravity in the
Randall-Sundrum background with matter on the brane is formulated. 
Using a novel gauge, in which the transverse components of the metric
are non-vanishing, the brane is kept straight. We analyze the
gauge symmetries and identify the physical degrees of freedom
of gravity. Our results underline the necessity for
non-gravitational confinement of matter to the brane.}

\newpage

\section{Introduction and Summary}\label{intro}
There has been much interest recently in the studies of non-compact
spaces containing 3-branes as domain walls as an alternative to
compactification for the treatment of the hierarchy problem
\cite{Arkani98,Antoniadis98,Arkani00,Randall99a,Cohen99}. 
These scenarios, in fact, revive older ideas described in
\cite{Rubakov83,Akama83,Visser85}. 
In a simple, non-compact scenario, which was proposed by Randall and
Sundrum (RS) \cite{Randall99b}, the 3-brane is flat due to fine tuning
of either the brane tension or the cosmological constant. Gravity on
the brane resembles the usual 4-dimensional Einstein gravity for long
distances, owing to a graviton bound state, whereas the
Kaluza Klein modes nearly decouple from the gravity on the brane
\cite{Lykken99,Brandhuber99}. 

One of the most interesting questions in this and similar scenarios is
the coupling of gravity to matter on the brane, because it represents
the very mechanism of generating gravitational forces.
In fact, several papers have studied this question, and the existence
of long-distance Einstein gravity on the brane has been confirmed
\nocite{Gogberashvili99,Halyo99,Chamblin99,deWolfe99,Shiromizu99,Garriga99,Youm99,Sasaki99,Ivanov99}
\nocite{Myung00-1,Dick00,Youm00a,Kallosh00,Myung00-2,Giddings00,Grojean00,Mueck00-1,Dvali00-1,Csaki00-1,Gregory00-2,Hawking00-2,Dvali00-2,Dadhich00-1,Kogan00-1,Csaki00-2,Gregory00-3,Emparan00-1,Kang00,Youm00c,Duff00-1,Kim00}
\cite{Gogberashvili99}--\cite{Kim00}.
Many of these calculations have used the RS gauge \cite{Randall99b},
as a result of which the brane appears bent owing to matter located on
it \cite{Garriga99,Myung00-2,Giddings00,Csaki00-1,Dvali00-2}. Moreover,
calculations in global Riemannian normal coordinates are simply not
consistent \cite{Mueck00-1}.

In this paper, we would like to follow up on the issue of localization
of gravity on the brane and address the following points: First, we
present a globally consistent formulation of linearized gravity with
matter located on the brane, which is kept \emph{straight}, 
both in the RS and an alternative background. Our motivation for
studying the alternative background is that the brane attracts
ordinary matter in that case, whereas it is repulsive in the
Randall-Sundrum case \cite{Mueck00-1}. Consistency of our calculation
is made possible by adopting a novel gauge fixing (cf.\
Secs.~\ref{method} and \ref{calc}). In our gauge, the metric
components transversal to the brane are non-vanishing. Fluctuations of
this kind have been studied already in \cite{Ivanov99}. 
Second, we determine the physical degrees of freedom of gravity by
analyzing the gauge freedom for a straight brane (Sec.~\ref{gauge}).
The problem of physical degrees of freedom was studied in
\cite{Myung00-2,Dvali00-2,Kang00}. 
Last, we find the particular
solution for a matter perturbation on the brane and analyze the
effective 4-dimensional gravity (Sec.~\ref{gravity}). 
We would like to mention that our solution is only gauge-equivalent to
the bent-brane formulation \cite{Garriga99,Giddings00}, if we are far
enough from the brane, although also the results for gravity on the
brane seem to be identical. Further comments can be found in
Sec.~\ref{gauge}. 

Although our
results are not entirely new, in that they confirm the
well-established long-distance Einstein gravity, they represent, to
our mind, a concise and elegant global formulation and provide important
insights into the dynamics on the brane. In particular, a
non-gravitational mechanism of confinement for the RS
background turns out to be essential
not only because of the instability of the geodesics on the brane 
\cite{Mueck00-1,Gregory00-3}, but also for the Newtonian dynamics on
the brane, which would be spoiled by extrinsic contributions. 
On the other hand, Einstein's equation does not hold for long distances
for the alternative background.

Let us now summarize our results. Our metric has the form 
\begin{equation}
\label{i:metric}
  ds^2 = \e{-2k|y|} (\eta_{\mu\nu} +\gamma_{\mu\nu}) dx^\mu dx^\nu +
  2n_\mu dx^\mu dy + (1+\phi) dy^2,
\end{equation}
where $\eta_{\mu\nu}=\textrm{diag}(-1,1,1,1)$. In eqn.\
\eqref{i:metric} and henceforth, positive $k$, $k>0$, corresponds to the RS
background and negative $k$, $k<0$, to the alternative background.  

Using the more general ansatz \eqref{i:metric} instead of the RS gauge
for the metric we are able to keep the brane straight at $y=0$ and
obtain a globally consistent solution for the linearized Einstein
equations. For a matter perturbation $t_{\mu\nu}$ located on the brane, our
solutions for the first order quantities in eqn.\ \eqref{i:metric} are  
\begin{equation}
\label{i:nmuphi}
  n_\mu = - \frac{\sgn{y}}{8k} \gamma_{,\mu}, \quad
  \phi = - \frac{\sgn{y}}{4k} \gamma_{,y},
\end{equation}
and the traceless transversal part of $\gamma_{\mu\nu}$
satisfies
\begin{multline}
\label{i:dgl}
  \partial_y \left( \e{-2k|y|} \partial_y \tilde\gamma_{\mu\nu}
  \right) - 2k\sgn{y} \e{-2k|y|} \partial_y \tilde\gamma_{\mu\nu}
  + \Box \tilde\gamma_{\mu\nu} \\
  = -16\pi \delta(y) \left[ t_{\mu\nu}
  - \frac13 \left( \eta_{\mu\nu} - \frac{\partial_\nu
  \partial_\mu}{\Box}\right)t \right].   
\end{multline}
Moreover, the trace $\gamma$ satisfies the boundary conditions
\[ \gamma|_{y=0} =  \frac{32\pi k}{3\Box} t, \quad
\gamma|_{y=\infty} =0, \]
but can be altered by gauge transformations for $y\ne0$. 
The step functions in eqn.\ \eqref{i:nmuphi} correspond to an apparent
singularity separating two coordinate patches. We shall demonstrate
explicitely in Sec.~\ref{gauge} how to obtain a continuous solution. 

\section{General Method}\label{method}
The starting point for our calculation is the action
\begin{equation}
\label{action}
  S = \int d^4 x \int dy \sqrt{-g} (R-2\Lambda) + \int_{\text{brane}} d^4x
  \sqrt{-\hat g} (\sigma +\mathcal{L}_{\text{matter}}),
\end{equation}
where 
\begin{equation}
\label{finetune}
  \Lambda = -6 k^2 \quad \text{and} \quad \sigma = - 12k,
\end{equation}
so that the metric
\begin{equation}
\label{bmetric}
  ds^2 = \e{-2k|y|} \eta_{\mu\nu} dx^\mu dx^\nu +dy^2
\end{equation} 
is a background solutions for $\mathcal{L}_{\text{matter}}=0$. 

In our approach, we would like to use a global coordinate system, in which the
brane is located at $y=0$ even when a perturbation is present on the
brane. Thus, it is natural to use the time-slicing formalism
\cite{MTW}, where we slice with respect to $y=const.$
hypersurfaces. At least in the RS case ($k>0$), we cannot impose an a 
priori gauge on the metric perturbations, since we would obtain
exponentially growing solutions for large $y$. Instead, consistency 
forces us to use a very particular, yet elegant, gauge. 
This shall become clear in the course
of our calculation. Let us now give a short review of the time-slicing
formalism \cite{MTW} and briefly outline the specific character of our
approach. 

In the time slicing formalism, one splits up the metric tensor as 
\begin{equation}
\label{metric1}
 \left( g_{ab} \right) = \begin{pmatrix}  
   \hat g_{\mu\nu} &n_\mu\\ n_\nu & n_\lambda n^\lambda + n^2 
 \end{pmatrix}, \qquad
 \left( g^{ab} \right) = \frac1{n^2} \begin{pmatrix}  
   n^2 \hat g^{\mu\nu} + n^\mu n^\nu & - n^\mu\\ - n^\nu & 1
 \end{pmatrix},
\end{equation}
where $\hat g_{\mu\nu}$ is the induced metric on the hypersurfaces,
$n^\mu = \hat g^{\mu\nu} n_\nu$, and $a,b=0,1,2,3,5$,
$\mu,\nu=0,1,2,3$. Henceforth, we shall denote quantities derived from
the induced metric $\hat g_{\mu\nu}$ with a hat. $n$ and $n_\mu$ are
the lapse function and shift vector, respectively. 

We consider the hypersurfaces $y=const.$, which have the induced
metric $\hat g_{\mu\nu}$ and the tangent and normal vectors
\begin{gather}
\label{tangent} 
  \partial_\mu x^a = \begin{cases} \delta_\mu^a \quad &\text{for
  $a=0,1,2,3$,}\\ 0 &\text{for $a=5$,}\end{cases} \\
\intertext{and} 
\label{normal}
  N_a = (0,0,0,0,-n), \qquad N^a = \frac1n (n^\mu,-1),
\end{gather}
respectively. Then, the second fundamental form measuring the
extrinsic curvature of the hypersurfaces has the form
\begin{equation} 
\label{H}
  H_{\mu\nu} = \frac1{2n} \left( \partial_y \hat g_{\mu\nu} - 
  \hat\nabla_\mu n_\nu - \hat\nabla_\nu n_\mu \right).
\end{equation}

Einstein's equation,
\begin{equation}
\label{einst1}
  R^{ab} -\frac12 g^{ab} R = - g^{ab} \Lambda + 8\pi T^{ab},
\end{equation}
can now be rewritten in terms of $\hat g_{\mu\nu}$, $n_\mu$ and
$n^2$. First, using the Gauss-Codazzi equations, the normal and
mixed components of eqn.\ \eqref{einst1} with respect to the
hypersurfaces become
\begin{align} 
\label{einst2}
  \hat R + H^\mu_\nu H^\nu_\mu -H^2 &= 2\Lambda,\\
\label{einst3}
  \partial_\mu H - \hat\nabla_\nu H^\nu_\mu &= 0,
\end{align}
where we have used the fact that $N_a T^{ab}=0$ for the energy
momentum tensor derived from the action \eqref{action}. For the
tangential components of eqn.\ \eqref{einst1} we prefer the form  
\begin{equation}
\label{einst4}
  R_{\mu\nu} = \frac23 \hat g_{\mu\nu} \Lambda + 8\pi 
  \left( T_{\mu\nu} -\frac13 \hat g_{\mu\nu} T \right), 
\end{equation}
because it saves us from calculating the scalar curvature $R$.
In the standard approach, one fixes $n_\mu$ and $n^2$ to some
convenient value. Then, eqn.\ \eqref{einst4} is the equation of motion
for $\hat g_{\mu\nu}$, whereas eqns.\ \eqref{einst2} and
\eqref{einst3} are constraints. Notice that until now
all expressions have been exact.

In a previous paper \cite{Mueck00-1}, some of us followed the standard
approach choosing globally $n_\mu=n^2-1=0$ and found that the linearized
approximation was inconsistent for the Randall-Sundrum
background. Therefore, we would now like to 
take a different approach. Instead of fixing the lapse function $n$
and the shift vector $n_\mu$ a priori, we leave them present at first
and fix them in the course of our calculations by the condition that
the linearization be consistent. For the linearization 
of the Randall-Sundrum and the alternative background, we write the
induced metric as 
\begin{equation}
\label{ghat}
  \hat g_{\mu\nu} = \e{-2k|y|} \left( \eta_{\mu\nu}+
  \gamma_{\mu\nu} \right), 
\end{equation}
and we consider $\gamma_{\mu\nu}$, $n_\mu$ and $n^2-1$ as small
perturbations. Furthermore, we assume that the induced metric
perturbations, $\gamma_{\mu\nu}$, are continuous at $y=0$.
The necessary linearized expressions for the connections and
curvatures are given in the appendix.

\section{Linearized Equations and Gauge Choice}\label{calc}
Let us start by linearizing Einstein's equations. 
The energy momentum tensor, as found from the action \eqref{action},
has the form 
\begin{equation}
\label{tmunu}
  T^{\mu\nu} = -\frac{3k}{4\pi} \sqrt{\frac{\hat g}{g}}\, 
  \delta(y) \hat g^{\mu\nu}   
  + \delta(y) t^{\mu\nu}(x), \quad T^{5\mu}=T^{55} =0,
\end{equation}
where the first term of $T^{\mu\nu}$ is the background from the brane, and
$t^{\mu\nu}$ is a small matter perturbation sitting on the brane.
The covariant conservation law, $\nabla_a T^{ab}=0$ is
satisfied to first order, if and only if $t^{\mu\nu}$ is conserved in
the conventional sense, $\partial_\mu t^{\mu\nu}=0$.

The constraints, eqns.\ \eqref{einst2} and \eqref{einst3}, take the
linearized forms 
\begin{gather}
\label{einst22}
  \e{2k|y|} \left( \gamma^{\mu\nu}{}_{,\mu\nu} -\Box \gamma - 6k
  \sgn{y} \partial^\mu n_\mu\right) = - 3k\sgn{y} \gamma_{,y} -12
  k^2 \phi\\
\intertext{and}
\label{einst32}
  \e{2k|y|}\left( \Box n_\mu - \partial_\mu \partial^\nu n_\nu
  \right) = - 3k\sgn{y} \partial_\mu \phi + 
  \partial_y \left( \gamma^\nu{}_{\mu,\nu} - \gamma_{,\mu} \right),
\end{gather}
respectively, where we have defined 
\begin{equation}
\label{phidef}
  \phi = n^2-1.
\end{equation}

Next, let us linearize the tangential equation, eqn.\ \eqref{einst4},
with the energy momentum tensor \eqref{tmunu}. We
have $\hat g/g\approx 1-\phi$, so that one finds the linearized form 
\begin{multline}
\label{einst42}
  \frac12 \left(\gamma^\rho{}_{\mu,\rho\nu} +
  \gamma^\rho{}_{\nu,\rho\mu} - \Box \gamma_{\mu\nu} -
  \gamma_{,\mu\nu} \right)
  + 3k^2 \phi \e{-2k|y|}\eta_{\mu\nu}
  -\frac12 \partial_y \left(\e{-2k|y|} \gamma_{\mu\nu,y} \right)\\ 
  -\frac12 \partial_\mu \partial_\nu \phi
  + \frac12 \partial_y (n_{\mu,\nu} + n_{\nu,\mu})
  - k\sgn{y} \left( n_{\mu,\nu} + n_{\nu,\mu} + \eta_{\mu\nu} 
  \partial^\lambda n_\lambda \right)\\  
  + k \sgn{y} \e{-2k|y|} \left[ \gamma_{\mu\nu,y} 
  + \frac12 \eta_{\mu\nu} \gamma_{,y} \right] 
  - \frac12 k \eta_{\mu\nu} \partial_y \left( \e{-2k|y|}\sgn{y}
  \phi\right) \\ 
  = 8\pi \delta(y) \left(t_{\mu\nu} -\frac13 \eta_{\mu\nu} t \right),
\end{multline}
where $t=\eta^{\mu\nu} t_{\mu\nu}$.

In the Randall-Sundrum case, the exponentially
increasing terms in eqns.\ \eqref{einst22} and \eqref{einst32} are
potentially problematic, but problems can be avoided by choosing a suitable
gauge. Thus, instead of setting $n_\mu=\phi=0$, we impose the gauge
conditions 
\begin{align}
\label{gauge1}
  4k\phi &= - \sgn{y} \gamma_{,y},\\
\label{gauge2}
  \partial^\mu \tilde\gamma_{\mu\nu} &=0,\\
\intertext{where}
\notag 
  \tilde\gamma_{\mu\nu} &= \gamma_{\mu\nu} - \frac14 \eta_{\mu\nu}
  \gamma  
\end{align}
is the traceless part of $\gamma_{\mu\nu}$. Eqn.\ \eqref{gauge2} says that
the traceless part $\tilde\gamma_{\mu\nu}$ should also be
transversal. 

Together, the gauge conditions \eqref{gauge1} and \eqref{gauge2} imply
that the right hand sides of eqns.\ \eqref{einst22} and
\eqref{einst32} are zero, so that these equations reduce to 
\begin{align}
\label{nmu1}
  \gamma^{\mu\nu}{}_{,\mu\nu} -\Box \gamma &=  6k \sgn{y} 
  \partial^\mu n_\mu, \\
\label{nmu2}
  \Box n_\mu - \partial_\mu \partial^\nu n_\nu &= 0, 
\end{align}
respectively. Eqns.\ \eqref{nmu1} and \eqref{nmu2} are equations for
the shift vector $n_\mu$, and their general solution is 
\begin{equation}
\label{nmu}
  n_\mu = - \frac{\sgn{y}}{8k} \gamma_{,\mu} + A_\mu.
\end{equation}
Here, the vector $A_\mu$ satisfies the 4-dimensional equations of a
free vector field in Lorentz gauge,
\begin{equation}
\label{Amu}
  \Box A_\mu = \partial^\mu A_\mu =0,
\end{equation}
but it can depend also on $y$.

After inserting the gauge conditions \eqref{gauge1} and \eqref{gauge2}
as well as the solution \eqref{nmu} for the shift vectors into eqn.\
\eqref{einst42}, we find 
\begin{multline}
\label{eqmot1}
  \partial_y \left( \e{-2k|y|} \partial_y \tilde\gamma_{\mu\nu}
  \right) - 2k \sgn{y} \e{-2k|y|} \partial_y
  \tilde\gamma_{\mu\nu} + \Box \tilde\gamma_{\mu\nu} \\
  - \partial_y (A_{\mu,\nu} + A_{\nu,\mu})  
  + 2k \sgn{y} (A_{\mu,\nu} + A_{\nu,\mu}) \\
  = \delta(y) \left[ -16 \pi \left(t_{\mu\nu} -\frac13 \eta_{\mu\nu} t
  \right) - \frac1{2k} \gamma_{,\mu\nu} \right].
\end{multline}

Notice that by virtue of eqns.\ \eqref{gauge2} and \eqref{Amu} the
left hand side of eqn.\ \eqref{eqmot1} is traceless and
four-divergence-free. Thus, we should expect the same of the right
hand side. The former property translates into 
\begin{equation}
\label{boxgamma}
  \Box \gamma|_{y=0} = \frac{32\pi k}3 t,
\end{equation}
whereas the latter property is expressed as 
\begin{equation}
\label{bcdiv}
  -16\pi \left( \partial^\mu t_{\mu\nu} -\frac13 \partial_\nu t
  \right) - \frac1{2k} \partial_\nu \Box \gamma|_{y=0} =0. 
\end{equation}
Using eqn.\ \eqref{boxgamma} we find $\partial^\mu t_{\mu\nu}=0$,
which is a good check of concistency, since Einstein's equation should
imply the covariant energy-momentum conservation law. 

Looking at the equations presented so far we realize that there is
no equation of motion for $\gamma$, and eqns.\ \eqref{Amu} and
\eqref{eqmot1} are insufficient to determine $A_\mu$ and
$\tilde\gamma_{\mu\nu}$. Therefore, we suspect that there is residual
gauge freedom, the determination of which is the subject of the next
section. 

\section{Physical Degrees of Freedom}
\label{gauge}
We would like to discuss the residual gauge freedom left after
imposing the gauge conditions \eqref{gauge1} and \eqref{gauge2} and
keeping the brane fixed at $y=0$. We shall find that traceless
transversal spin-2 excitations are the only physical degrees of
freedem. Moreover, we shall show how to remove the step functions in
the solutions for $n_\mu$ and $\phi$ [cf.\ eqns.\ \eqref{gauge1} and
\eqref{nmu}]. 

To start, consider two
coordinate systems with the metrics 
\begin{equation}
\label{mettrafo}
\begin{aligned} 
  ds^2 &= \e{-2k|y|} \left(\eta_{\mu\nu} + \gamma_{\mu\nu} \right)
  dx^\mu dx^\nu + 2n_\mu dx^\mu dy + (1+\phi) (dy)^2\\
  &= \e{-2k|y'|} \left(\eta_{\mu\nu} + \gamma'_{\mu\nu} \right)
  dx'{}^\mu dx'{}^\nu + 2n'_\mu dx'{}^\mu dy' + (1+\phi') (dy')^2,
\end{aligned}
\end{equation}
related by an infinitesimal coordinate transformation,
\begin{equation}
\label{trafo}
  x'{}^\mu =x^\mu - \xi^\mu(x,y),\qquad y' = y -\xi^5(x,y).
\end{equation}
The location of the brane remains unchanged, i.e.\ we have the
restriction $\xi^5(x,0)=0$. This also ensures that no normal
components of the energy momentum tensor, $T^{a5}$, are generated. 

Under the coordinate transformation \eqref{trafo} the first order
elements transform as
\begin{align}
\label{gammatrafo}
  \gamma_{\mu\nu} &= \gamma'_{\mu\nu} + 2k \sgn{y} \xi^5 \eta_{\mu\nu} -
  \xi_{\mu,\nu} - \xi_{\nu,\mu},\\
\label{ntrafo} 
  n_\mu &= n'_\mu - \xi^5_{,\mu} - \e{-2ky} \xi_{\mu,y},\\
\label{phitrafo} 
  \phi &= \phi' -2\xi^5_{,y}.
\end{align}
Here, indices have been lowered using the Minkowski metric. 
With some calculations one can check that the linearized Einstein
equations, eqns.\ \eqref{einst22}, \eqref{einst32} and
\eqref{einst42}, are invariant under
this transformation, provided that $\xi^5|_{y=0}=0$.

Coordinate
transformations can be applied separately for $y<0$ and
$y>0$. For simplicity, we shall restrict our discussion to $y>0$.
For $y>0$, the trace of eqn.\ \eqref{gammatrafo} yields
\begin{equation}
\label{tracetrafo}
  \gamma = \gamma' + 8k \xi^5 - 2 \partial^\lambda \xi_\lambda,
\end{equation}
which can be combined with eqn.\ \eqref{gammatrafo} to obtain the
transformation of $\tilde\gamma_{\mu\nu}$,
\begin{equation}
\label{tildetrafo}
  \tilde\gamma_{\mu\nu} = \tilde\gamma'_{\mu\nu} 
  - \xi_{\mu,\nu} - \xi_{\nu,\mu} 
  + \frac12 \eta_{\mu\nu}\partial^\lambda \xi_\lambda.
\end{equation}
Thus, imposing the gauge condition \eqref{gauge2} on both metrics
yields 
\begin{equation}
\label{ximu}
  \Box \xi_\mu + \frac12 \partial_\mu \partial^\nu \xi_\nu =0.
\end{equation}

Next, we substitute the gauge condition \eqref{gauge1} into eqn.\ 
\eqref{phitrafo} and obtain $\partial_y \partial^\lambda
\xi_\lambda=0$, i.e.\ 
\begin{equation}
\label{f}
   \partial^\lambda \xi_\lambda = f(x) 
\end{equation}
is a function of $x$ only. Moreover, from eqn.\ \eqref{ximu} we find
that $f$ must satisfy $\Box f=0$. Finally, substituting the solution
for $n_\mu$, eqn.\ \eqref{nmu}, into eqn.\ \eqref{ntrafo}, one finds
that $A_\mu$ transforms as 
\begin{equation}
\label{strafo} 
  A_\mu = A'_\mu - \frac1{4k} \partial_\mu f(x) - \e{-2ky}
  \xi_{\mu,y}.
\end{equation}
As a check of consistency, we observe that eqn.\
\eqref{Amu} and the boundary condition \eqref{boxgamma} remain
unchanged. 

Let us now fix the remaining gauge freedom and identify the
physical degrees of freedom. First, we can use 
\[ \gamma|_{y=0} = \gamma'|_{y=0} - 2f(x) \] to obtain a \emph{unique}
solution for the boundary condition $\eqref{boxgamma}$, which we shall
formally write as 
\begin{equation}
\label{dirbcgamma}
  \gamma|_{y=0} = \frac{32\pi k}{3\Box} t.
\end{equation}
Second, we make use of $\xi^5$ 
in order to choose a convenient function for $\gamma$ in the bulk 
satisfying the boundary conditions \eqref{dirbcgamma}. 
For example, one could pick
\begin{equation}
\label{gammachoice}
  \gamma = \frac{32\pi k}{3\Box} t\, \e{-ay^2}
\end{equation}
with some positive coefficient $a$. Notice that on the brane
$\gamma$ is determined by the matter content, which cannot be gauged
away. Last, we make use of the remaining freedom, $\xi_\mu$ satisfying
$\partial^\mu \xi_\mu =\Box \xi_\mu =0$, and 
\[ A_\mu = A'_\mu - \e{-2ky} \xi_{\mu,y} \]
in order to pick a convenient function for $A_\mu$. Of course, the
most convenient value is $A_\mu=0$, which we shall adopt.

After this gauge fixing, we are left with only the physical degrees of
freedom $\tilde\gamma_{\mu\nu}$, which describe spin-2 gravity
excitations. 

We would like to point out the following subtle point regarding the
bent-brane formulation used by Garriga and Tanaka \cite{Garriga99},
Giddings, Katz and Randall \cite{Giddings00} and others.
If one did not impose the condition $\xi^5|_{y=0}=0$ [cf.\ eqn.\
\eqref{trafo}], it would seem
from eqns.\ \eqref{gammatrafo}--\eqref{phitrafo} that one could
transform a metric in our gauge into a metric in Randall-Sundrum gauge
using $\xi^5=-\gamma'/8k$. The obvious effect would be that the
brane appears bent to an observer, and non-zero normal
components $T^{5\mu}$ are generated. However, we would like to emphasize
that eqn.\ \eqref{mettrafo} is not valid in this case, because one
cannot expand $\e{-2k|y'+\xi^5|}\approx\e{-2k|y'|}(1-2k\sgn{y}\xi^5)$, 
which would imply that the brane is located again
at $y'=0$ (the brane is where the singularity of the curvature is). 
Rather, one should write
\[ ds^2 = \e{-2k|y'+\xi^5|}(\eta_{\mu\nu} + \gamma''_{\mu\nu})
{dx'}^\mu {dx'}^\nu + {dy'}^2. \]
Thus, eqn.\ (10) of \cite{Garriga99} should contain
$\delta[y-\xi^5(x)]$ and not $\delta(y)$.
In our opinion, this seems to be a drawback of the bent-brane
formulation and has not been addressed properly. 

As a last point in this section, we would like to demonstrate how to
remove the step functions in the solutions for $n_\nu$ and $\phi$
[cf.\ eqn.\ \eqref{i:nmuphi}] and to obtain a continuous solution. 
First let us note that it is enough to remove the
discontinuity of $n_\mu$, because one can choose $\gamma$ such that
$\phi|_{y=0}=0$ [cf.\ eqn.\ \eqref{gammachoice}], i.e.\ $\phi$ is already
continuous. We take a gauge transformation of the form 
\begin{equation}
  \xi^5=0, \quad 
  \xi_\mu= \frac1{16k^2} \gamma_{,\mu} \e{2k|y|} \psi(y),
\end{equation}
where $\psi(y)$ is a smooth function with a compact support such that
$\psi(0)=1$. Then, the gauge-transformed components of the metric will
be 
\begin{align}
\label{gammaprime}
  \gamma'_{\mu\nu} &= \gamma_{\mu\nu} +\frac1{8k^2} \e{2k|y|} \psi(y)
  \gamma_{,\mu\nu}, \\
\label{nprime}
  n'_\mu &= -\frac{\sgn{y}}{8k} \gamma_{,\mu} + \frac1{16k^2}
  \e{-2k|y|} \partial_y \left[ \gamma_{,\mu} \e{2k|y|} \psi(y)
  \right],\\
\label{phiprime}
  \phi' &= -\frac{\sgn{y}}{4k} \gamma_{,y}.
\end{align}
One can easily see that $n'_\mu$ is continuous at $y=0$. 
Thus, we have found an explicit variable transformation for a
vicinity of the brane, which transforms our solution into one that is
continuous at $y=0$. We can, therefore, conclude that the step
functions in eqn.\ \eqref{i:nmuphi} correspond only to an apparent
singularity.

\section{Particular Solution and Gravity on the Brane}\label{gravity}
Let us now fix the gauge as described in the last section 
and solve eqn.\ \eqref{eqmot1}. Substituting $A_\mu=0$ and 
eqn.\ \eqref{dirbcgamma}, eqn.\ \eqref{eqmot1}
becomes
\begin{multline}
\label{eqmot1a}
  \partial_y \left( \e{-2k|y|} \partial_y \tilde\gamma_{\mu\nu}
  \right) - 2k\sgn{y} \e{-2k|y|} \partial_y \tilde\gamma_{\mu\nu}
  + \Box \tilde\gamma_{\mu\nu} \\
  = -16\pi \delta(y) \left[ t_{\mu\nu}
  - \frac13 \left( \eta_{\mu\nu} - \frac{\partial_\nu
  \partial_\mu}{\Box}\right)t \right].
\end{multline}
We note that the inverse of the d'Alembertian is unique after the
residual gauge fixing. 

The particular solutions for the source $t_{\mu\nu}$ are even in $y$. 
Therefore, we shall consider eqn.\ \eqref{eqmot1a} for $y>0$,
\begin{equation}
\label{eqmot2}
  \partial_y \left( \e{-2ky} \partial_y \tilde\gamma_{\mu\nu}
  \right) - 2k \e{-2ky} \partial_y \tilde\gamma_{\mu\nu}
  + \Box \tilde\gamma_{\mu\nu} =0
\end{equation}
and impose the Neumann boundary condition
\begin{equation}
\label{bc} 
  \partial_y \tilde\gamma_{\mu\nu}|_{y=+0} = -8\pi \left[ t_{\mu\nu}
  - \frac13 \left( \eta_{\mu\nu} - \frac{\partial_\nu
  \partial_\mu}{\Box}\right)t \right],
\end{equation}
which arises from integrating eqn.\ \eqref{eqmot1a} over the
singularity.

Fourier transforming eqn.\ \eqref{eqmot2} with respect to the brane
coordinates and changing variables to $z=\e{2ky}$ leads to the
differential equation
\begin{equation}
\label{eqmot3}
  \left( z^2 \partial_z^2 - z\partial_z - \frac{p^2}{4k^2} z \right)
  \tilde\gamma_{\mu\nu} =0,
\end{equation}
whose solutions are given in terms of Bessel functions
\cite{Gradshteyn}. 

Let us consider the case $p^2>0$. The static case
($p_0=0$) is included here, and the case $p^2=0$ can be obtained as a
limit from $p^2>0$. The two linearly independent solutions of
eqn.\ \eqref{eqmot3} are  
\begin{equation}
\label{sol1}
  \tilde\gamma_{\mu\nu}(p,y) = c_{\mu\nu}(p) \e{2ky} \begin{cases} 
  \K_2\left(\e{ky} |p| /k \right) \\
  \I_2\left(\e{ky} |p| /k \right) \end{cases} \quad (p^2>0).
\end{equation}
Since blowing-up solutions are inconsistent with the linearization
(and clearly unphysical), we are led to choose the solution with the
$\K$ function for $k>0$ (Randall-Sundrum background) and the
solution with the $\I$ function for $k<0$ (alternative
background). Then, we can determine the coefficients $c_{\mu\nu}(p)$
from the boundary condition \eqref{bc}. After doing so, the final
solution for $\tilde\gamma_{\mu\nu}$ reads  
\begin{equation}
\label{solution}
  \tilde\gamma_{\mu\nu}(p,y) = \frac{8\pi}{|p|} \left[ t_{\mu\nu}
  - \frac13\left( \eta_{\mu\nu} -\frac{p_\mu p_\nu}{p^2}
  \right) t\right] \e{2k|y|} \begin{cases} 
  \frac{\K_2\left(\e{k|y|} |p|/k \right)}{\K_1(|p|/k)}
  \quad&\text{(RS),}\\ 
  \frac{\I_2\left(\e{k|y|} |p|/k \right)}{\I_1(|p|/k)}.
  \end{cases}
\end{equation}

We can use the solution \eqref{solution} to discuss the effective laws
of gravity on the brane. First, we would like to know the
gravitational potential due to, but far away from a static point
source, which we introduce by 
\[ t_{00}(x) = M/\mpl{5}^3 \delta(\vec{x}),\qquad 
t_{00}(p) = 2\pi \delta(p_0) M/\mpl{5}^3,\]
where $\mpl{5}$ is the Planck mass in five dimensions. 

There are two possibilities for the behaviour of a test particle
on the brane. First, one could assume that the particle is free to
move in five dimensions, i.e.\ it will follow a geodesic in
five-space, which for small velocities is 
\begin{equation}
\label{geod1}
  \frac{d^2 x^i}{dt^2} \approx - \Gamma^i{}_{00}
\end{equation}
for $i=1,2,3$. Then, 
\begin{equation} 
\label{Gammai00} 
  \Gamma^i{}_{00} = \hat\Gamma^i{}_{00} + k\sgn{y} n_i = -\frac12
  \partial_i \gamma_{00} - \frac18 \partial_i \gamma = -\frac12
  \partial_i \tilde\gamma_{00},
\end{equation}
where we have used eqn.\ \eqref{nmu} and the fact that our particular 
solution is static. Since we are interested in the long distance
behaviour on the brane, we set $y=0$ in eqn.\ \eqref{solution} and
consider small $|p|$. Using $\K_2(z) \approx 2/z^2$, $\K_1(z) \approx
1/z$, $\I_2(z) \approx z^2/8$ and $\I_1(z)\approx z/2$, we find 
\[ \tilde\gamma_{00}(p,0) \approx 
 \begin{cases} \frac{32\pi k}{3p^2} t_{00} +\cdots \quad &\text{(RS)},\\
 \frac{4\pi}{3k} t_{00} +\cdots . \end{cases} \]
The $1/p^2$ term for the Randall-Sundrum background 
generates a $1/r$ potential term about the static point source,
whereas there is no $1/r$ term in the alternative case.
More specifically, we can deduce from eqns.\ \eqref{geod1} and
\eqref{Gammai00} that  
\begin{equation}
\label{newton1}
  V_{\text{unconstr}}(r) = \begin{cases} 
  -\frac{4kM}{3\mpl{5}^3r} +\cdots \quad &\text{(RS),}\\
  \text{no $1/r$ term}  
  \quad&\text{(alternative)}. \end{cases}
\end{equation}

The second possibility is that the particle is constrained to move along
the brane by some non-gravitational mechanism. This would mean that it
follows a geodesic on the brane, i.e.
\begin{equation}
\label{geod2}
  \frac{d^2 x^i}{dt^2} \approx - \hat\Gamma^i{}_{00}.
\end{equation}
However, 
\begin{equation} 
\label{hatGammai00} 
  \hat\Gamma^i{}_{00} = -\frac12 \partial_i \gamma_{00} = 
  -\frac12 \partial_i \tilde\gamma_{00}+ \frac18 \partial_i \gamma,
\end{equation}
where $\gamma$ is given by the Dirichlet boundary condition
\eqref{dirbcgamma}. Hence, the gravitational potential is 
\begin{equation}
\label{newton2}
  V_{\text{constr}}(r) = V_{\text{unconstr}}(r) - \frac{4\pi k}{3\Box}
  t_{00} \approx \begin{cases} 
  -\frac{kM}{\mpl{5}^3r} +\cdots \quad &\text{(RS),}\\
  -\frac{kM}{3\mpl{5}^3r} +\cdots &\text{(alternative)}. \end{cases}
\end{equation}

Obviously, in the unconstrained case the dynamics on the brane is
affected by the non-zero shift 
vectors. For the RS background, the freedom of
a test-particle to move in five dimensions would imply that the
trajectories on the brane are unstable, since the brane is repulsive
\cite{Mueck00-1}. Hence, we are lead to favour the situation, in which
the test particle is confined to move along the brane by some
non-gravitational mechanism. Then, potential shortcuts via the fifth 
dimension are not allowed, the dynamics is determined by
the intrinsic metric only, and eqn.\ \eqref{newton2} applies. 
Moreover, in the alternative background,
only confinement to the brane leads to a $1/r$ potential.

Let us also comment on the solution of eqn.\ \eqref{eqmot3} for
$p^2<0$, i.e.\ for the case of tachyonic matter sources on the brane.
In that case, the linearly independent solutions of eqn.\
\eqref{eqmot3} are
\begin{equation}
\label{sol2}
  \tilde\gamma_{\mu\nu}(p,y) = c_{\mu\nu}(p) \e{2ky} \begin{cases} 
  \N_2\left(\e{ky} |p| /k \right) \\
  \J_2\left(\e{ky} |p| /k \right) \end{cases} \quad (p^2<0).
\end{equation}
In the Randall-Sundrum background, both modes diverge for large
$y$. (Although the Bessel functions both go to zero, the exponential
factor in front diverges faster.) Interestingly, these modes are
integrable, since the norm integral contains a factor $\e{-4k|y|}$
in the invariant integration measure cancelling the diverging factor;
and they describe the massive Kaluza Klein modes used to construct the
five dimensional Green's function in \cite{Garriga99,Giddings00}. In
fact, we have performed the integral over the Kaluza Klein states for
$p^2>0$ in the Green's function \cite[eqn.\ (13)]{Garriga99} and found
perfect agreement with our solution \eqref{solution}. 
However, classical solutions containing the $p^2<0$ modes will diverge
for $y\to\infty$. Therefore, in the RS scenario,
the existence of tachyonic matter on the brane is inconsistent with 
linearized gravity in the five-dimensional space-time, as is the
existence of free Kaluza Klein gravity excitations. The non-existence of both
might be quite desirable in a theory describing the real world.

As a second objective we would like to consider the zero-mode
truncation of the solution \eqref{solution} on the brane. The
intrinsic Einstein tensor on the brane can be written as 
\begin{equation}
\label{et}
  \hat R_{\mu\nu} -\frac12 \eta_{\mu\nu} \hat R = -\frac12 \Box \tilde
  \gamma_{\mu\nu} - \frac14 \left( \gamma_{,\mu\nu} - \eta_{\mu\nu}
  \Box \gamma \right),
\end{equation}
where we have used the gauge \eqref{gauge2}. In the
Randall-Sundrum case, using only the
zero-mode of eqn.\ \eqref{solution}, we find 
\begin{equation}
\label{zeromode}
  \Box \tilde \gamma_{\mu\nu} \overset{\text{zero-mode}}{=} -16\pi k 
  \left[ t_{\mu\nu} - \frac13 \left(\eta_{\mu\nu}
  -\frac{\partial_\mu\partial_\nu}{\Box} \right) t\right].
\end{equation}
Furthermore, we can substitute the Dirichlet boundary condition
\eqref{dirbcgamma} for $\gamma$, so that eqn.\ \eqref{et} becomes 
\begin{equation}
  \hat R_{\mu\nu} -\frac12 \eta_{\mu\nu} \hat R
  \overset{\text{zero-mode}}{=} 8\pi k t_{\mu\nu}.
\end{equation}
The same equation can be obtained using $\gamma'_{\mu\nu}$, which is
related to $\gamma_{\mu\nu}$ by a four-dimensional gauge
transformation [cf.\ eqn.\ \eqref{gammaprime}].
Thus, the zero-mode truncation of the solution \eqref{solution} yields
Einstein's equation on the brane for the Randall-Sundrum background,
which by now is a well-established result
(cf.\ e.g.\ \cite{Randall99b,Garriga99,Dick00}).

It is easy to see that this is not the case for the alternative background,
as in that case eqn.\ \eqref{zeromode} would not be valid.   

\section{Conclusions}\label{concl}
In this paper, we have used a novel gauge in order to obtain
a solution of the linearized Einstein equations in the
Randall-Sundrum and an alternative background, where the brane is kept
straight in spite of matter perturbations located on it. Our solution
is consistent in each of the two half-spaces $y>0$ and $y<0$, and the
two patches can be connected by making the gauge transformation
\eqref{gammaprime}--\eqref{phiprime}. 

The explicit solution was summarized already by the equations \eqref{i:metric},
\eqref{i:nmuphi} and \eqref{i:dgl}, and a particular solution of
eqn.\ \eqref{i:dgl} is found in eqn.\ \eqref{solution}.
Our analysis of the gauge degrees of freedom showed that the traceless
transversal part of $\gamma_{\mu\nu}$, 
$\tilde\gamma_{\mu\nu}$, represents all physical degrees of freedom. 
In particular, we conclude that the unphysical graviscalar mode
mentioned in \cite{Dvali00-2} is a gauge mode. This was found
independently in \cite{Kang00}. 

Based on our solution, we studied the effective laws of gravity on the
brane and found, in the Randall-Sundrum background, that the zero-mode
truncation yields Einstein's equation for the intrinsic metric on the
brane. This implies the validity of the Newtonian limit, if the
dynamics is determined by intrinsic quantities on the brane only, and
agrees with our derivation of the Newton potential. Moreover, it
emphasizes the importance of non-gravitational confinement of matter
to the brane in order to prevent the extrinsic geometry from entering
the dynamics of matter on the brane. 
A non-gravitational confinement is also necessary
for the geodesics along the brane to be stable \cite{Mueck00-1,Gregory00-3}.
In this article, we did not discuss the corrections to the Newton 
potential on the brane, as this has been done elsewhere 
\cite{Randall99b,Garriga99,Mueck00-1,Duff00-1}.

In this work, we have restricted our discussion to thin branes. It
would be interesting to study the analogue for thick branes, which
would also provide an appropriate regularization.

\section*{Acknowledgments}
We would like to thank R. Brandenberger, G. Dvali, S. Giddings,
L. Randall, K. Sfetsos, R. Sundrum and T. Tanaka for fruitful
discussions and correspondence. 
I.~A. and I.~V. are grateful to the Physics Department of Simon
Fraser University for its kind hospitality. 
Partial support of this research came from the following grants: 
RFFI 99-01-00166 and INTAS 99-0545 for I.~A., the grant for leading
scientific schools 96-15-96208 for M.~I. and I.~V., NSERC for W.~M. and
K.~V., and RFFI 99-01-00105 and INTAS 99-0590 for I.~V.

\section*{Appendix}
Here, we list various linearized expressions necessary for the
calculations in the main text. The metric tensor has the form
\eqref{metric1}, where the induced metric $\hat g_{\mu\nu}$ is
linearized by eqn.\ \eqref{ghat}, and $\gamma_{\mu\nu}$, $n_\mu$ and
$n^2-1$ are small perturbations.
We shall henceforth raise and lower the indices of $\gamma_{\mu\nu}$
and of $\partial_\mu$ (and only of these) with the
Lorentz metric. 

First, the connection coefficients intrinsic to the hypersurfaces are 
\begin{align*} 
  \hat\Gamma^\mu{}_{\nu\lambda} &= \frac12 \left( 
  \gamma^\mu{}_{\nu,\lambda} + \gamma^\mu{}_{\lambda,\nu} - 
  \gamma_{\nu\lambda}{}^{,\mu} \right),\\
\intertext{and the intrinsic Ricci tensor and curvature scalar are,
  respectively}
  \hat R_{\nu\rho} &= \frac12 \left( \gamma^\mu{}_{\nu,\mu\rho} + 
  \gamma^\mu{}_{\rho,\mu\nu} - \Box \gamma_{\nu\rho} -
  \gamma_{,\nu\rho} \right),\\
  \hat R &= \e{2k|y|} \left( \gamma^{\mu\nu}{}_{,\mu\nu} - \Box
  \gamma \right).
\end{align*} 

The linearized expression for the second fundamental
form, (cf.\ eqn.\ \eqref{H}) is 
  \[ H^\mu_\nu = \frac1{2n} \left[-2k\sgn{y} \delta^\mu_\nu + 
  \gamma^\mu{}_{\nu,y} - \e{2k|y|} \eta^{\mu\lambda} \left(
  n_{\nu,\lambda} + n_{\lambda,\nu} \right) \right]. \]

The necessary connection coefficients of the five-space are 
\begin{align*}
  \Gamma^\mu{}_{\nu\lambda} &= \hat\Gamma^\mu{}_{\nu\lambda} -
  k\sgn{y} \eta_{\nu\lambda} \eta^{\mu\rho} n_\rho,\\
  \Gamma^y{}_{\nu\lambda} &= \frac1{n^2}\left(k\sgn{y} \hat
  g_{\nu\lambda} \right) + \frac12 \left( n_{\nu,\lambda} +
  n_{\lambda,\nu} - \e{-2k|y|} \gamma_{\nu\lambda,y} \right),\\
  \Gamma^\mu{}_{\nu y} &= - k\sgn{y} \delta^\mu_\nu + \frac12 
  \left[ \gamma^\mu{}_{\nu,y} + \e{2k|y|} \eta^{\mu\lambda} 
  \left(n_{\lambda,\nu} - n_{\nu,\lambda} \right) \right],\\
  \Gamma^y{}_{\nu y} &= k\sgn{y} n_\nu + \frac12 (n^2-1)_{,\nu},\\
  \Gamma^y{}_{yy} &= \frac12 (n^2-1)_{,y}.
\end{align*}
Thus, the linearized Ricci tensor becomes
\begin{align*}
  R_{\mu\nu} &= \hat R_{\mu\nu} - \frac{4k^2}{n^2} \hat g_{\mu\nu} 
  + \frac{2k}{n^2} \delta(y) \hat g_{\mu\nu} 
  - \frac12 \left(\e{-2k|y|} \gamma_{\mu\nu,y}\right)_{,y}
  - \frac12 (n^2-1)_{,\mu\nu}\\
  &\quad + \frac12 (n_{\mu,\nu}+n_{\nu,\mu})_{,y}
  - k\sgn{y} \left( n_{\mu,\nu}+n_{\nu,\mu} +\eta_{\mu\nu} 
  n_\lambda{}^{,\lambda} \right)\\
  &\quad + k\sgn{y} \e{-2k|y|} \left\{ \gamma_{\mu\nu,y} + \frac12
  \eta_{\mu\nu} \left[\gamma_{,y} - (n^2-1)_{,y} \right]\right\}.
\end{align*}

\end{document}